\documentclass[sn-mathphys-num]{sn-jnl}

\usepackage{graphicx}
\usepackage{multirow}
\usepackage{amsmath,amssymb,amsfonts}
\usepackage{amsthm}
\usepackage{mathrsfs}
\usepackage[title]{appendix}
\usepackage{xcolor}
\usepackage{textcomp}
\usepackage{manyfoot}
\usepackage{booktabs}
\usepackage{algorithm}
\usepackage{algorithmicx}
\usepackage{algpseudocode}
\usepackage{listings}

\theoremstyle{thmstyleone}

\theoremstyle{thmstyletwo}

\theoremstyle{thmstylethree}

\begin{document}

\title[Article Title]{Solar Cycles: Can They Be Predicted?}

\author{Floe Foxon}

\affil{University of Leeds, Leeds, United Kingdom}

\abstract{The solar magnetic field, thought to be generated by the motion of plasma within the Sun, alternates on the order of 11-year cycles and is incompletely understood. Industries rely on accurate forecasts of solar activity, but can solar cycles be predicted? Of more than 100 predictions for cycle 25, most underestimated the amplitude (peak sunspot number). Fewer predictions were made for the timing of solar maximum, but timing predictions seem to be performing better than amplitude predictions. Reasons for inaccurate prediction are suggested, and perspectives are given on how future studies might improve upon the extant literature.}


\maketitle

\section*{Introduction} \label{sec:intro}

The last solar cycle, number 24, was the weakest in a century. The current cycle 25 (C25) was originally predicted as another weak cycle, and was still described as ``weak'' through 2023 \citep{Barbuzano}, but 2024 saw a burst of solar activity, pushing C25 beyond C24's activity. C25 is expected to peak in 2025 (and may have already peaked in 2024), but the timing will not be clear until months after. Given the importance of assessing the performance of previous forecasts to direct future studies, here I summarize over 100 predictions for C25.

Predictions rely on indicators such as sunspot numbers (SSNs; Fig.~\ref{fig1}), which have been observed for thousands of years, and which vary over the course of a cycle; increasing toward solar maximum (peak solar activity) and decreasing toward solar minimum. Some predictions have relied on other solar measurements, such as the solar radio flux; radio emissions from the Sun at a wavelength of 10.7 cm (F10.7) measured since the last century.

\section*{Cycle 25 predictions}

\begin{figure}[ht!]
\centering
\includegraphics[width=1.0\textwidth]{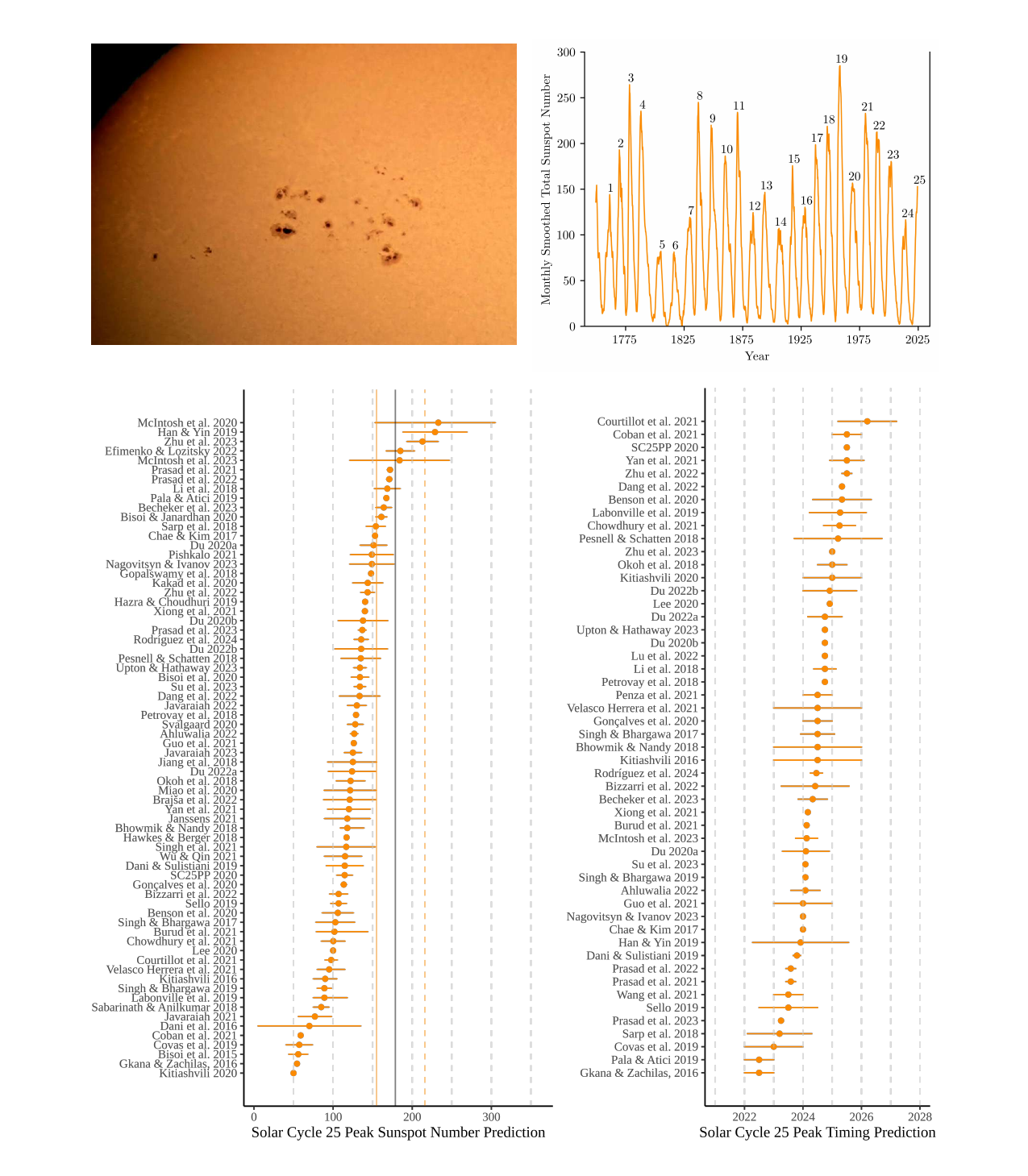}
\caption{(Upper Left) Sunspot region cluster near solar maximum. (Upper Right) Smoothed monthly total SSN from 1749CE--present. Data from: \url{https://www.sidc.be/SILSO/datafiles}. (Lower Left) Published predictions for amplitude of C25, excluding SSN series v1.0-based predictions. Solid orange vertical line: highest observed smoothed SSN as of this writing (July 2024). Dashed orange vertical line: highest observed raw SSN as of this writing (August 2024). Solid black vertical line: average peak SSN across cycles 1--24 (historical baseline). (Lower Right) Published predictions for timing of C25's peak solar activity. Images by Floe Foxon.}\label{fig1}
\end{figure}

The Astrophysics Data System and others return more than a hundred C25 predictions published between 1983 and 2024, many published after 2015 (Fig.~\ref{fig1}). Excluding predictions based on the obsolete SSN v1.0 series, the average prediction for the amplitude of C25 was 127 units of SSN, ranging from 50--233.

As of this writing, SSN data are available through January 2025. So far, the highest smoothed value is 154.9 (July 2024), and the highest raw value for a given month is 216 (August 2024); these already exceed the average prediction for C25. Only a handful of predictions gave maximum sunspot numbers consistent with the August 2024 peak. The average amplitude of cycles 1--24 was 178.7; this historical baseline is closer to the observed values than most C25 predictions.

Simply put, most predictions underestimated C25, perhaps because of C24's weakness. Oppositely, many C24 predictions \textit{over}estimated that cycle's amplitude, perhaps because of how strong cycles 21--23 were \citep{Nandy}.

Fewer predictions were made for the timing of solar maximum, but these are apparently performing better than amplitude predictions. Although the actual timing of solar maximum is unknown as of this writing, it is now thought to occur some time in 2025 (or peaked already in 2024); many predictions did include 2024/5 in their ranges.

\section*{Discussion}

Although it is tempting to think that the existence of an eleven-year cycle and longer-term trends in solar activity make the solar cycle predictable, there is an element of apparent randomness in solar cycles; cycles are not all the same duration, peaks do not occur at the same point for each cycle, cycles can have multiple peaks, and the gradient of ascent of SSN varies considerably. Additionally, most models are based on the `smooth' SSN averaged across months of observation; this makes sense from a modelling perspective, but activity can vary dramatically from month-to-month, and smoothed-data models cannot predict this variability.

Solar activity may be semi-unpredictable due to (a) hidden/inaccessible variables, or (b) patterns yet to be observed since SSNs have only been counted for around 300 years (with variable reliability), while solar cycles have likely occurred for billions of years. Either of the above would confound long-term solar cycle prediction.

Now that C25 is peaking, predictions for cycle 26 are already being made \citep{Rodriguez}. Solar cycles may yet be semi-predictable, but many C25 prediction methods relied on simple linear correlations between solar parameters, without necessarily demonstrating predictive validity. What's more, newer machine learning techniques did not generally outperform older methods, e.g. several neural network/deep learning models predicted that C25's amplitude would be similar to or weaker than that of C24 \citep{Okoh, Benson, Bizzarri, Covas}. Some predictions did not include estimates of uncertainty, and some that did were unrealistically low.

Future predictions may be aided by further revisions to the SILSO SSN series, and by further work on historical solar activity data from terrestrial measurements such as isotope abundances in tree rings and polar icecaps (e.g. \cite{Brehm}).

Of course, it is possible for a model to `accurately' predict the current cycle, but fail to predict future cycles; the true measure of a model is in its ability to successfully predict future cycles time and again. Current methods struggle to model the next single solar cycle, much less the next 15 cycles \citep{Hiremath}. It may be that only late-stage predictions are possible; the few amplitude predictions that were consistent with the August 2024 peak were all published between 2019 and 2023. This comports with \citet{Werner}'s conclusion during the previous cycle that ``accurate forecasting over a period of time longer than 2 to 4 years seems impossible.''

It remains to be seen whether the few modeling methods that did not underestimate the activity of C25 will predict the activity of cycles 26 and on with similar accuracy. To do so would be profound, as they may help describe the magnetic nature of the Sun and other main sequence stars. Indeed, \citet{Schussler} concluded that while not all models are just ``numerology or parameter-tuning... the key point is not so much to predict but to understand the solar cycle.'' As \citet{Pesnell} put it: ``Solar cycle predictions are an interesting and practical area of research. Only by making and analyzing these predictions will we build accurate models of the solar dynamo.''

\section*{Acknowledgements}

Based on work presented at the 245th AAS meeting: \url{https://doi.org/10.17605/OSF.IO/QGJRA}.

\bibliography{FoxonBib}

\end{document}